\documentclass[10pt, conference]{IEEEtran}
\IEEEoverridecommandlockouts

\usepackage{subfigure}

\usepackage[utf8]{inputenc}
\usepackage{amsmath,amssymb}

\usepackage{graphicx}
\usepackage{subfigure}
\usepackage{color}
\usepackage{array}
\usepackage{url}

\usepackage{algorithm}
\usepackage{algorithmic}

\usepackage[colorinlistoftodos,prependcaption,textsize=small]{todonotes}
\usepackage{blindtext}
\presetkeys%
    {todonotes}%
    {inline,backgroundcolor=yellow}{}

\begin{document}

\title{Simulation of the Internet of Things\footnotemark}

\author{\IEEEauthorblockN{Gabriele D'Angelo, Stefano Ferretti, Vittorio Ghini}
\IEEEauthorblockA{Department of Computer Science and Engineering, University of Bologna\\
Bologna, Italy\\
\{g.dangelo, s.ferretti, vittorio.ghini\}@unibo.it}
}


\maketitle

\footnotetext{The publisher version of this paper is available at \url{http://dx.doi.org/10.1109/HPCSim.2016.7568309}.
\textbf{{\color{red}Please cite this paper as: ``Gabriele D'Angelo, Stefano Ferretti, Vittorio Ghini. Simulation of the Internet of Things. Proceedings of the IEEE 2016 International Conference on High Performance Computing and Simulation (HPCS 2016)''.}}}

\begin{abstract}
This paper presents main concepts and issues concerned with the simulation of Internet of Things (IoT).
The heterogeneity of possible scenarios, arising from the massive deployment of an enormous amount of sensors and devices, imposes the use of sophisticated modeling and simulation techniques.
In fact, the simulation of IoT introduces several issues from both quantitative and qualitative aspects.
We discuss novel simulation techniques to enhance scalability and to permit the real-time execution of massively populated IoT environments (e.g., large-scale smart cities).
In particular, we claim that agent-based, adaptive Parallel and Distributed Simulation (PADS) approaches are needed, together with multi-level simulation, which provide means to perform highly detailed simulations, on demand. We present a use case concerned with the simulation of smart territories.
\end{abstract}


\section{Introduction}
\label{sec:intro}

An unprecedented number of connected devices will soon be added to the Internet. A multitude of sensors and mobile users' terminals are designed to interact in order to offer novel services in smart cities and territories in general.
These devices, in the so-called Internet of Things (IoT), have very specific characteristics both in terms of hardware (in many cases, these devices are equipped with a very little amount of memory and computational power), software (specific OSes) and management (little or no administration utilities, few system updates). Being able to understand and to simulate the IoT will soon become essential. The complex networks obtained by the interaction of IoT devices are hard to design and to manage. In real deployment scenarios, many possible configurations of IoT networks are possible. Devices’ connectivity is influenced by their geographical location, communication and network capabilities, device distribution.

The modeling of a general IoT environment to build effective and smart services can be quite difficult, due to the heterogeneous possible scenarios. Thus, IoT simulation is necessary for both quantitative and qualitative aspects. To name a few issues: capacity planning, what-if simulation and analysis, proactive management and support for many specific security-related evaluations. The scale of the IoT is the main problem in the usage of existing simulation tools. Traditional approaches (that are single CPU-based) are often unable to scale to the number of nodes (and level of detail) required by the IoT. 

The main goal of this paper is to introduce the main aspects of the simulation of IoT, discussing a new combination of techniques to enhance scalability and to permit the real-time execution of massively populated IoT environments (e.g., large-scale smart cities). For example: parallel and distributed simulation, adaptive computational and communication load-balancing, self-clustering, multi-level modeling and simulation. 

To demonstrate the validity of the proposed approach, an application scenario of ``smart shires'' is analyzed~\cite{smartshires,smartshires_abps}.
This is a novel view of devising smart, cheap and sustainable services in decentralized geographical spaces, without the need of costly (communication) infrastructures. Such services would make good use of a deployment of cheap sensors in these areas, together with ad-hoc configurations of mobile devices.
We show that the design and configuration of smart services in (decentralized) territories impose the simulation of wide area networks; however, in certain cases a highly detailed simulation is required. This need for scalability and high level of detail can be reached only through properly configured multi-level simulation techniques.
An advantage of this approach is that the detailed (and thus, more costly) simulation can be performed only when needed, in a limited simulated area, only for the needed time interval of the simulation.\\

The remainder of this paper is organized as follows. Section \ref{sec:background} describes the background about IoT/Smart-Territories and the Simulation Techniques. In Section \ref{sec:stateoftheart} the state of the art related to IoT simulation is discussed. The proposed approach, based on adaptive parallel/distributed simulation and multi-level simulation, is introduced in Section \ref{sec:multilevelsimulation}. In Section \ref{sec:casestudy}, this approach is applied to a ``smart shires'' case study. Finally, Section \ref{sec:conc} provides some concluding remarks.

\section{Background}
\label{sec:background}

\subsection{Internet of Things and Smart-Territories}

As already mentioned, there is an important trend towards the design of novel services, built by interconnecting various heterogeneous devices deployed in geographical areas \cite{Petrolo:2014}. Data sensed by the sensors' devices can be disseminated and collected by some information processing system, treated as open data and managed through a context-aware data distribution service, to be used by applications \cite{iot_survey}. 

Sensors are relatively cheap in terms of costs. Thus, their massive deployment is feasible both in populated centres and in more decentralized areas \cite{smartshires}. Such sensors can be interconnected to form a sensor network. In turn, such information gathered through this network can be passed to services placed within cloud (or fog) computing architectures. These smart services can integrate such data with crowd-sensed and crowd-sourced data coming from mobile terminals. 
The approach of exploiting any kind of information coming from the cloud of things available in the territory is now referred as Sensing as a Service (SaaS) model \cite{Khan:2012,Lea:2014,PereraZCG14}. 

The complexity of the possible scenarios coming from this picture suggests that effective simulation tools are needed. These simulation tools must take into consideration issues concerned with complex networks, aspects typical of pervasive computing, low-level details concerned with wireless communications. In the next sections, we will discuss on existing methods and viable strategies to ensure scalability of the simulation, without introducing oversimplifications and inaccuracies, due to the lack of the level of detail.

\subsection{Discrete Event Simulation}
In a computer simulation, a process models the behavior of some other system over time~\cite{FUJ00}. In some cases, the simulated system is real but more often it has yet to be designed or implemented. In practice, simulation is about methodologies and techniques that are needed for the performance evaluation of complex systems.

The motivations behind the use of simulation are many. To name a few: cost reasons, testing on the real system is too dangerous, many different solutions must to be evaluated to support the system design (i.e.~dimensioning and tuning). Due to the increasing complexity in the systems to be built, simulation is used more and more often. 

Discrete Event Simulation (DES)~\cite{Law:1999:SMA:554952} is one of the many simulation paradigms that have been proposed. With respect to other approaches, it is has good expressiveness and it is quite easy to use. A DES is represented by a simulated model (that is implemented using a set of state variables) and its evolution (that is represented by a sequence of events processed in chronological order). Each event occurs at a given instant in time and represents a change in the simulated model state. This means that the whole evolution of the simulated system is obtained through the execution of an ordered sequence of events that are: created, stored and processed. For example, the events in the simulation of Vehicular Ad Hoc Networks are the updates of the cars positions and the transmission of data packets. At the basics, a DES is a set of state variables (i.e.~describing the modeled system), an event list (i.e.~the pending events that will be processed for evolving the simulated state) and global clock (i.e.~the simulation time)~\cite{Law:1999:SMA:554952}. Each event is tagged by a timestamp that specifies the simulated time at which it has to be processed.

In a sequential (i.e.~monolithic) simulation, a single Physical Execution Unit (PEU), for example a CPU core, is in charge of creating new events, updating the pending event list and processing the events in timestamp order. In other words, a single CPU core manages the whole simulated model and its evolution. This approach is simple and easy to implement but it has some drawbacks. Among others, the simulation scalability both in terms of execution time (to complete the simulations runs) and size of the system that can be represented~\cite{1668384}.

\subsection{Parallel DES and PADS}
As an alternative, the tasks described above can be parallelized using a set of interconnected PEUs (e.g.~CPU cores, CPUs or hosts). This approach is called Parallel Discrete Event Simulation (PDES)~\cite{Fujimoto:1989:PDE:76738.76741}. In this case, very large and complex models can be represented and executed since each PEU is only in charge of a part of the simulation model. That is, each PEU manages a local pending events list and some events are delivered by means of messages to remote PEUs. In addition, the PEUs must run a synchronization algorithm to guarantee the correct simulation execution. In many cases, a PDES approach can speedup the simulation execution, at the cost of a more complex implementation and setup of the simulator.

A Parallel and Distributed Simulation (PADS) is a simulation that is run on more than one processor~\cite{perumalla2007}. There many good reasons to rely on this approach, among them: execution speed, model scalability, interoperability and composability purposes (e.g.~to integrate different off-the-shelf simulators and to compose many already existing simulation models in a new simulator)~\cite{FUJ00}.

\begin{figure}[ht]
\centering
\includegraphics[width=7.0cm]{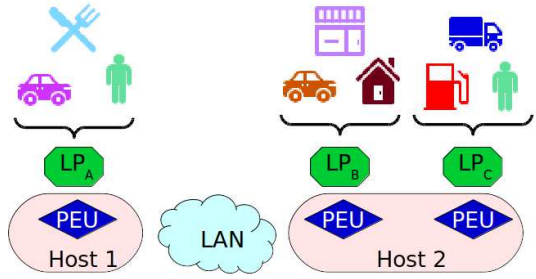}
\caption{Parallel and Distributed Simulation: model partitioning.}
\label{fig:pads-model}
\end{figure}

With respect to a monolithic simulation, a PADS lacks of a global model state. That is, a single representation of the simulated model is missing. In fact, each PEU in the PADS manages only a part of the simulated model. Following the PADS terminology, the model components executed on top of each PEU are called Logical Processes (LPs)~\cite{gda-hpcs-11}. As shown in Figure~\ref{fig:pads-model}, a PADS is obtained through the interaction among LPs, each one of which, deals with the evolution of a part of the simulated model and interacts with the other LPs (for synchronization and data distribution)~\cite{FUJ00}.

The performance of the network that interconnects the LPs has a strong effect on the PADS design and the simulator execution speed. When the LPs are run on PEUs interconnected by a shared memory, then it is called parallel simulation. Conversely, loosely coupled LPs are referred as distributed simulation. More often, the execution architecture used to run PADS are a mix of parallel and distributed PEUs~\cite{gda-simpat-2014}.

In short, the main issues in a PADS are:
\begin{itemize}
	\item the simulated model is partitioned in a set of LPs~\cite{bagrodia98}. The \textbf{partitioning} is a complex task given that it must be done considering both the minimization of the network communication (among LPs) and the load balancing in the parallel/distributed execution architecture;
	\item the results obtained by the PADS are correct only if they are exactly the same given by the sequential simulator.	This can happen only if there is a \textbf{synchronization} algorithm that properly coordinates the LPs evolution;
	\item each LP generates updates (events) that are possibly relevant for parts of the simulated model in other LPs. For performance reasons, broadcasting all events is not feasible. \textbf{Data distribution} is about the efficient delivery of state updates and it is often based on a publish-subscribe approach~\cite{Jun:2002:ESM:564062.564074}.
\end{itemize}

Implementing a PDES using PADS is obtained encapsulating the events in messages for their inter-LP delivery. As defined by Lamport: ``two events are in causal order if one of them can have some consequences on the other''~\cite{Lamport1978}. Clearly, to get a correct simulation execution, the causal order of events must not being violated. This is easy in a monolithic simulation but it is complex in parallel and distributed architectures due to the different execution speeds of each PEU and the network delays. In a PADS, to guarantee that all events are executed in in non-decreasing timestamp order, the LPs have to run a synchronization algorithm. The synchronization can be handled in many ways but the main approaches are the following:
\begin{itemize}
	\item \emph{time-stepped}: the simulated time is divided in timesteps	of fixed-size. The simulation model is updated at every timestep and the lower bound to the flight time for interactions between the model components is the size of the timestep. When a LP completes the tasks for the current timestep, it broadcasts to all the other LPs an End-Of-Step (EOS) message and then waits the EOS messages from all other LPs before proceeding to the next timestep~\cite{1261535};
	\item \emph{conservative}: in this approach the causality errors are prevented. That is, before processing each event, it is checked if the event is ``safe'' or not (with respect to the causality constraint). If the event is tagged as safe by the synchronization algorithm then it can be processed. Otherwise, the LP must stop processing while waiting for more events (or better information about the safety of events). This safety check can be implemented in many different ways, a widely used is algorithm is the Chandy-Misra-Briant~\cite{misra86};
	\item \emph{optimistic}: in this case the events are processed by the LPs in receiving order. This means that, very likely, the causality order will be violated. In fact, when a violation is found by the synchronization algorithm, the LP that has found it implements a roll-back to the (most recent) previous state that is correct. Furthermore, it propagates the roll-back to all the other LPs that have been affected by the violation~\cite{timewarp}. In this way, the whole PADS goes back to the most recent globally correct simulation state and it starts again processing the events.
\end{itemize}

\subsection{Adaptive PADS}
Ad described before, in PADS the partitioning of the simulated model is a complex task. In~\cite{gda-simpat-2014}, we have proposed an approach in which the simulated model is represented by a multi-agent system. The simulated model is partitioned in small model components (also said Simulated Entities, SEs) and the model evolution is obtained through the exchange of interactions among SEs. In this way, the LPs are containers of SEs and it is possible to move (migrate) a SE from a LP to another. This permits to avoid the static partitioning of the simulated model and to adaptively reallocate the SEs for better computational and communication load balancing. In many cases, this lead to a speed up in the simulation execution and enhanced scalability.

This adaptive PADS approach is implemented in the GAIA/ART\`IS simulator~\cite{pads} and, in the current version, it is based on a time-stepped synchronization. That, as described in the follow of this paper, is also at the basis of the multi-level modeling that we propose for the simulation of IoT models.

\section{State of the Art}
\label{sec:stateoftheart}

\subsection{Simulation of the Internet of Things}

The design of complex IoT setups requires the support of large scale testbeds or the usage of scalable simulation tools. In the case of simulation, the number of nodes in the scenario and the level of detail required by the interaction among nodes are key elements for the scalability of the simulator.

In~\cite{6069710}, the authors identify the requirements for the next generation of IoT experimental facilities, they discuss some drawbacks of simulation-based approaches and provide a survey of existing testbeds (some of them also supporting co-simulation). An approach based on the federation of testbeds is possible but it has many drawbacks. In many cases, an on simulation would be preferred but the existing network simulators are inadequate for the scale and level of detail required by IoT models.

SimIoT is a new simulator described in~\cite{6844677} in which the back-end operations are executed in a cloud environment for better performance. The use case proposed in the paper is a health monitoring system for emergency situations in which short range and wireless communication devices are used to monitor the health of patients. The preliminary performance evaluation is based on 160 identical jobs submitted by 16 IoT devices.

In~\cite{6664581} the massive-scale of many IoT deployments is considered. In this case, the authors firstly present a survey of large-scale simulators and emulators and then they propose MAMMotH, a software architecture based on emulation. To the best of our knowledge the development of MAMMotH has stopped in 2013.

Brambilla et al. propose to integrate the DEUS general-purpose discrete event simulation with the domain specific simulators Cooja and ns-3 for the study of large-scale IoT scenarios in urban environments~\cite{Brambilla:2014:SPL:2694768.2694780}. In this case, the performance evaluation is based on 6 scenarios with up to $200000$ sensors, $400$ hubs and $25000$ vehicles. The execution time with respect to the number of events shows a quite good scalability. On the other hand, to the best of own knowledge, the DEUS simulator has a monolithic architecture and it is implemented in Java.

In~\cite{6418824} the authors propose an IoT-based smart home system in which the performance evaluation is based on different simulation methods such as Monte Carlo.

DPWSim is a simulation toolkit that supports the modeling of the OASIS standard ``Devices Profile for Web Services'' (DPWS)~\cite{6803226}. Its main goal of is to provide a cross-platform and easy-to-use assessment of DPWS devices and protocols. In other words, it is not designed for very large-scale setups.

The approach followed in \cite{Brumbulli2016} is to use a model-driven simulation (based on the standard language SDL) to describe the IoT scenario.  Starting from this, an automatic code generation transforms the description into an executable simulation model for the ns-3 network simulator.

Finally, an interesting approach is proposed in \cite{paper+kirsche-13:iot-simulation}. The author proposes a hybrid simulation environment in which the Cooja-based simulations (i.e.~system level) are integrated with a domain specific network simulator (i.e.~OMNeT++).

\subsection{Internet of Things and Smart-Territories}

As concerns the use of IoT to build efficient services for making ``smarter'' territories, from a simulation point of view
there are many requirements that the simulation tool must provide. Above all, the main issue is scalability, both in terms of amount of modeled entities and granularity of events. Even a small size smart territory will be composed by thousands of interconnected devices. Many of them will be mobile and each with very specific behavior and technical characteristics~\cite{smartshires}. 
If a proactive approach is needed (e.g.~simulation in the loop), in order to perform ``what-if analysis'' during the management of the deployed architecture, then the simulator should be able to run in (almost) real-time, at least with average size model instances. 

We claim that a multi-level simulation is needed in order to simulate a smart territory scenario with a reasonable IoT model. In fact, running the whole model at the highest level of detail is unfeasible. A better approach is to bind different simulators together, each one running at its appropriate level of detail and with specific characteristics of the domain to be simulated (e.g.~mobility models, wireless/wired communications and so on). We will discuss this approach in the next section.

Agent-based simulation is a perfect tool to create models that mimic urban systems in general~\cite{Karnouskos}. Agent-based simulation, together with land-use transport interaction model and cellular automata are applicable in planning support systems. These models can be applied at different time scales, such as short-term modeling, e.g. diurnal patterns in cities, and long-term models for exploring change through strategic planning.
Tools such as MASON \cite{Luke:2005} and SUMO \cite{SUMO2012} allow simulating moving entities (e.g.~mobile users of vehicles) that can interact with static ones. These tools have been successfully exploited to study intelligent traffic control systems \cite{bauza,kerekes,Wegener:2008,e16052384}, mobile applications that resort to crowdsensed data \cite{PrandiFMS15} and so on. The main problem of these approaches is that, due to their nature, they do not allow creating massive scenarios, with many interconnections.

CupCarbon is a multi-agent and discrete event, smart-city and Internet of Things Wireless Sensor Network (SCI-WSN) simulator \cite{Mehdi:2014}. Its allows designing, visualizing and validating distributed algorithms in a network. It employs the OpenStreetMap framework to deploy sensors directly on the map. The main goal of this tool is to help trainers to explain the basic concepts and how sensor networks work and it can help scientists to test their wireless topologies, protocols, etc. The main problem of scalability remains.

Moreover, it is worth mentioning that there is a number of image and 3D based simulators, such as CanVis, Second Life, Suicidator City Generator, Blended Cities. Among them, 
UrbanSim is a software-based simulation for urban areas, with tools for examining the interplay between land
use, transportation, and policy \cite{urbansim}. It is intended for use by Metropolitan Planning Organizations and others needing
to interface existing travel models with new land use forecasting and analysis capabilities.
UrbanSim does not focus on scenario development, as most of these tools do, but rather
on understanding the consequences of certain scenarios on urban communities.
However, typically such a kind of tools do not cope with issues concerned with wireless communications and pervasive computing, which are the keywords related to the IoT world.

\section{Multi-level Simulation}
\label{sec:multilevelsimulation}

Since many IoT models are composed of a very large number of nodes, the usage of fine grained simulation models leads to scalability problems in the performance evaluation. In other words, a monolithic simulator that handles all the nodes in the IoT and implements a fine grained level of detail is unable to provide the simulation results in an acceptable amount of time. Even using a PADS approach, the massively populated setups are difficult to handle. This can be overcome by: i) employing High Performance Computing execution platforms or ii) reducing the level of detail in the simulation model. Both these solutions are not feasible since the first is very costly and the second often leads to misleading (or wrong) simulation results due to the excessive amount of details removed from the simulated model.

\begin{figure*}[ht]
\centering
\includegraphics[width=.7\linewidth]{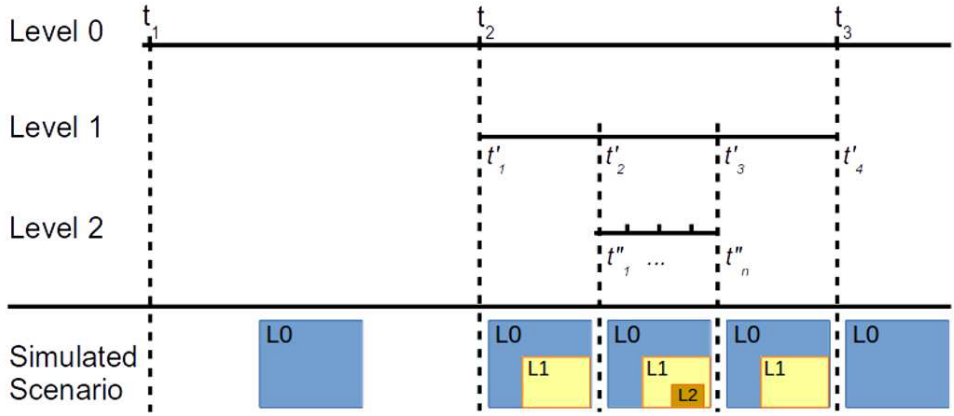}
\caption{Multi-level simulation.}
\label{fig:multilevel-simulation}
\end{figure*}

For these reasons, we propose a multi-level modeling and simulation~\cite{Ghosh:1986:CDM:319541.319559} approach for large scale IoT setups. That is, a simulation in which multiple simulation models are glued together~\cite{magne2000towards}. Each one with a specific task and working with at a different level of detail.

Under the implementation viewpoint, this means a ``high level'' adaptive PADS simulator (i.e.~GAIA/ART\`IS) that works at a coarse grained level of detail and that coordinates the execution of a set of domain specific ``middle'' or ``low level'' simulators that are used only when a fine grained level of detail is necessary (e.g.~OMNeT++~\cite{omnet}, ns-3~\cite{ns3}, SUMO~\cite{sumo}). The switch between coarse and fine grained models can be automatic or triggered by the simulation modeler. For example, if a given simulated area is populated by too many wireless devices then a detailed simulation model could assess network capacity or congestion problems. The main issues with this multi-level approach are the interoperability among the simulators and the design of the inter-model interactions (e.g.~synchronization and state exchanges at runtime between model components).

More specifically, as shown in Figure~\ref{fig:multilevel-simulation}, at the simulation bootstrap the whole scenario is executed at level 0 (that is, with a minimal details). Hence, the high level simulator (e.g.~GAIA/ART\`IS) manages the evolution of all the model components and their interactions following a time-stepped synchronization approach~\cite{gda-mswim-2004}. At timestep $t_2$, it is found that a part of the simulated scenario (for example a specific zone in the simulated area or a specific group of modeled nodes) has to be simulated with more details. This means that, in the figure, a part of the simulated area is still modeled at level 0 while a specific zone is now managed by the level 1 model. If necessary, in the following of the simulation, a specific area can be further detailed using a level 2 model (and simulator). To simplify this discussion, if we consider only two levels then that all the model components managed by the level 0 simulator are evolved using $t$-sized timesteps and all the others use $t'$-sized timesteps. Timestep $t_2$ (that is the same of $t'_1$ for level 1) is the moment in which a part of the model components is transferred from the coarse grained simulator to the finer one. In the following, the components at level 0 will jump from $t_2$ to $t_3$ while the components simulated at level 1 will be updated at $t'_2$, $t'_3$ and $t'_4$ (that is the same of $t_3$ for level 0) but since there is no more need for such a level of detail, all the components simulated at level 1 are transferred again to the level 0 simulator. Following the constraints imposed by the time-stepped synchronization algorithm, all the interactions among level 0 simulated components can happen at every coarse grained timestep while the interactions at level 1 can happen at every fine grained timestep. Finally, the interaction between components managed at different levels can happen only at the coarse grained timesteps. That is, when there is a match between the timesteps at the different levels.

Using this approach the total number of nodes handled by the simulator does not changes but the level of detail used in the simulation evolution is adapted to the needs of the simulation model at runtime. In other words, this means that the simulation model is not run at the lowest level of detail for the whole duration of the simulation. Hence, it is possible to obtain a better scalability with respect to traditional simulation (monolithic or PADS) approaches. On the other side, it is clear that the multi-level modeling (as as every kind of model approximation) introduces a some amount of error in every analysis. As in every simulation, appropriate verification and validation techniques need to be used.

At the time of writing, we are finalizing the design of the multi-level simulator and we are working on a prototype implementation~\cite{smartshires,smartshires_abps} that includes the case study described in the following section.

\section{Case Study}
\label{sec:casestudy}

As an application scenario, we consider a main use case concerned with the need to provide smart services to territories, being them cities or more decentralized areas. In particular, we focus on ``smart shires'', a novel view of decentralized geographical spaces able to manage resources (natural, human, equipment, buildings and infrastructure) in a way that is sustainable and not harmful to the environment~\cite{smartshires,smartshires_abps}. The idea is to create novel, smart and cheap services, easily deployable without the need of costly infrastructures, that would improve the life of citizens and tourists.

The need for cheap solutions forces the use of crowd-sensed and crowd-sourced data coming from the IoT. Sensors are relatively cheap in terms of costs. Thus, their deployment in a countryside is feasible. These sensors need to be interconnected through the use of smart communication approaches~\cite{Petrolo:2014}. Data sensed by the sensors' devices are ma\-na\-ged by a distributed information processing system, hence enabling a context-aware data distribution~\cite{Bellavista:2012}.

A wide range of application scenarios are possible, ranging from proximity-based applications (e.g.~proximity-based social networking, advertisements for by-passers, smart communication between vehicles, etc.), security and public safety support, services related to the production chain in rural environments (smart agriculture, smart animal farming), smart traffic management systems.

As a specific use-case example, recently the ``km 0'' phenomenon gained a lot of interests in Italian and European foodie circles. This abbreviation for ``zero kilometers'' signifies local, low impact primary food ingredients. The idea is to prioritize the use of local and seasonal foods, avoid the use of genetically modified organisms so as to improve the quality of provided products and promote sustainable cooking. In spite of the growing interest in local products, there are relatively few places where one can buy these products directly from the producer. Thus, customers have to look for specialized weekend famers's markets or for farm direct purchases. Customers might be single users, ethical purchasing groups, restaurant owners. And quite often, this products research reveals to be not a simple task for customers. Thus, smarter scenarios are possible.

Let imagine a service that allows consumers subscribing to the availability of a certain product. Upon availability of such a product by a producer (e.g.~the farmer), he can publish a notification, which can inform subscribers of product availability plus other related information such as, for instance, his presence in next, near markets or other possible purchasing opportunities.
In view of such details, the consumer can plan to visit the market (so as to have the opportunity to select the products directly), book some specific items, quantities and so on. So far so good, there is plenty of publish-subscribe mechanisms that might help these producer/consumer interactions in order to build smarter services. 
However, more sophisticated services are possible. The market could be crowded with several (apparently similar) producers, the customer might do not know the location, he might have some physical disabilities, and thus he might need to be guided to the exact location of the producer, that is dynamically determined (hence, without the possibility of knowing the position in advance).
Then, once there, he might be interested in finding other possible interesting products.

To cope with these issues, producers can provide information on the fly, thanks to proximity-based services that may guide customers in a smart and effective market tour.
Based on the available technologies of the market, such services can be deployed in different ways. For instance, if a wireless infrastructure is available, then all the communications can pass through this network. Otherwise, some ad-hoc solution should be dynamically built, with producers that exploit their smart devices (e.g.~smartphones) to build multihop wireless communication and information dissemination strategies~\cite{smartshires_abps}.
Moreover, in case of intermittent connections, seamless communication strategies should be employed, that for instance might employ multihoming \cite{Ferretti2016390}.
Being partly composed of advertisements, general information on the market, publish messages looking for their subscribers, such message dissemination might be viably performed using some kind of epidemic dissemination protocol over a dynamic, opportunistic ad-hoc overlay, used in conjunction with application filtering techniques \cite{simplex,Ferretti2013481,Wirtz:2014}.

\begin{figure}[ht]
\centering
\includegraphics[width=\linewidth]{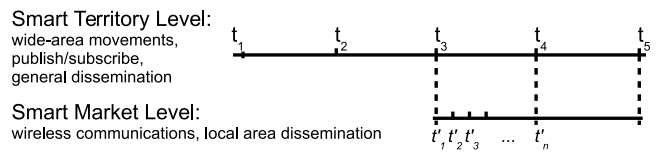}
\caption{Smart Territory/Market multilevel simulation.}
\label{fig:usecase-multilevel}
\end{figure}

The efficient simulation of such a wide scenario in a smart territory is not an easy task, since it involves several activities involving different domains and requiring very different levels of granularity. In this case, multi-level simulation can come into the picture. One can imagine different levels of granularity, as shown in Figure \ref{fig:usecase-multilevel}. The coarse level (level 0) simulates the whole smart territory, where different actors produce products, subscribe their interests, move towards different geographical areas. This can be implemented using some kind of classic agent-based simulator, maybe equipped with PADS capabilities \cite{gda-mospas-11}.

Then, once there is the need to simulate the specific interactions within the a specific area (e.g.~the ``smart market''), then more simulation details (and probably a different simulator) are needed to consider wireless communication issues, fine-grained interactions and movements. Thus, a more detailed level of simulation (based on a domain specific simulator) is triggered (i.e.~level 1 in the figure). In this case, each simulation step of the coarse grained simulation layer (e.g., $t_3, t_4$ of the level 0 in Figure~\ref{fig:usecase-multilevel}) is decomposed into multiple substeps at the fine grained layer (level 1). Following this approach, the level 1 simulator is able to notify level 0 with its simulation advancements.

\section{Conclusions}
\label{sec:conc}

In this paper, we discussed on main issues to cope with, in order to properly simulate the Internet of Things. Scalability and high level of details are the two main, and quite often counterposed, goals. We overviewed some existing techniques, reaching the conclusion that the use of adaptive, agent-based, Parallel and Distributed Simulation (PADS), coupled with multilevel simulation is a good strategy to employ in this context.

The analysis of the use case, related to the design of smart services for smart cities and decentralized areas, shows that multi-level simulation techniques provide means to simulate wide geographical areas, with a multitude of simulation entities (agents). However, when needed it is possible to trigger a more detailed, fine grained simulation, so as to consider aspects which could not be simulated otherwise. The interesting aspect of this approach is that the detailed (and more costly) simulation can be performed in a specific, limited simulated area, only for the needed time interval of the simulation.

\small{
\bibliographystyle{abbrv}
\bibliography{paper}  
}


\end{document}